# Optical-Clock-Based Time Scale

Jian Yao, Jeff A. Sherman, Tara Fortier, Holly Leopardi, Thomas Parker, William McGrew, Xiaogang Zhang, Daniele Nicolodi, Robert Fasano, Stefan Schäffer, Kyle Beloy, Joshua Savory, Stefania Romisch, Chris Oates, Scott Diddams, Andrew Ludlow, and Judah Levine

Time and Frequency Division, National Institute of Standards and Technology, Boulder, Colorado, USA, 80305

**Abstract**

A time scale is a procedure for accurately and continuously marking the passage of time. It is exemplified by Coordinated Universal Time (UTC), and provides the backbone for critical navigation tools such as the Global Positioning System (GPS). Present time scales employ microwave atomic clocks, whose attributes can be combined and averaged in a manner such that the composite is more stable, accurate, and reliable than the output of any individual clock. Over the past decade, clocks operating at optical frequencies have been introduced which are orders of magnitude more stable than any microwave clock. However, in spite of their great potential, these optical clocks cannot be operated continuously, which makes their use in a time scale problematic. In this paper, we report the development of a hybrid microwave-optical time scale, which only requires the optical clock to run intermittently while relying upon the ensemble of microwave clocks to serve as the flywheel oscillator. The benefit of using clock ensemble as the flywheel oscillator, instead of a single clock, can be understood by the Dick-effect limit. This time scale demonstrates for the first time sub-nanosecond accuracy for a few months, attaining a fractional frequency stability of $1.45\times10^{-16}$ at 30 days and reaching the $10^{-17}$ decade at 50 days, with respect to UTC. This time scale significantly improves the accuracy in timekeeping and could change the existing time-scale architectures.

## I. Introduction

Time is a dimension in which events can be ordered from the past through the present and into the future. Many modern-day technologies rely on the ability to do this accurately and precisely, including navigation [1], telecommunication systems [2], electrical power grids [3], and even electronic transactions on the stock exchange [4]. The most advanced timekeeping can be applied to fundamental science studies [5], such as searches for dark matter [6] and neutrino speed-measurements [7].

The microwave frequency of 9.192631770 GHz corresponding to the transition between the two hyperfine levels of the ground state of the Cesium atom has been used to define the SI second since 1967 [8], and such microwave clocks are the basis of international time. However, in actual timekeeping systems, an ensemble of atomic clocks based on Cesium, Hydrogen, and Rubidium is typically used. This forms a timescale with performance better than that afforded by an individual clock and also improves the reliability of the system [9]. Within this context, a new generation of atomic clocks, based on optical frequencies, have shown potential for tremendous improvement in timekeeping [10]. Examples of optical clock species include $Yb^+$ [11], $Al^+$ [12], Yb [13] and Sr [14], and the fractional frequency stability [*] of better than $1\times10^{-16}$ in just a few minutes or seconds has been demonstrated, representing orders-of-magnitude improvement over the best microwave clocks.

However, a key challenge is that these experimental optical clocks do not yet operate continuously for long intervals, making it difficult to incorporate them into conventional time scales. Recent efforts explore the combination of an intermittent optical clock with a continuous Hydrogen maser for time-scale generation [15-18]. As outlined in the green dashed box of Figure 1(a), the optical clock provides occasional frequency corrections to the Hydrogen maser to prevent the Hydrogen maser from deviating too far from the ideal time. In this "Hydrogen maser + optical clock" (HMOC) architecture, the performance is limited by the noise of the Hydrogen maser and the operation time of the optical clock. This intrinsic limitation has been explored theoretically for different conditions with numerical simulation [19], and can be further understood as an aliasing phenomenon of maser noise periodically sampled by the optical clock, referred to as the Dick effect (Appendix A).

This paper explores a novel way to improve an optical-clock-based time scale independent of the optical-clock operation time: improving the stability of the flywheel oscillator. We propose steering an ensemble of microwave clocks (e.g., a few Hydrogen masers), instead of a single Hydrogen maser, to an optical clock, (see the red dashed box of Figure 1(a)). Because the microwave time scale exhibits smaller noise than a single clock due to averaging, Dick-effect limitations in the steering process are reduced (Appendix A). This has also been confirmed by numerical simulation [19]. There, we showed that the stability of the optical-clock-based time scale is proportional to the square root of the number of Hydrogen masers. This "microwave time scale + optical clock" (MTSOC) architecture affords a time scale with improved performance at all averaging times, offering a complementary enhancement to that realized with increased optical-clock uptime. As an example, to achieve the performance of $4.0\times10^{-17}$ at $10^7$ sec (i.e., time deviation of 0.23 ns at $10^7$ sec), we can reduce the uptime of an optical clock from 50% to 8% by increasing the maser number from one to six (Figure 1(b)).

To test the MTSOC architecture, we conducted a campaign at NIST (National Institute of Standards and Technology, USA) from October 2017 to April 2018. Over the five months, the Yb clock operated intermittently, for an average of 1.5 hours per day. The free-running microwave time scale AT1, composed of a few Hydrogen masers and a few commercial Cesium clocks, is steered to the Yb clock using a Kalman filter. We show that the MTSOC architecture exhibits unprecedented timing accuracy (0.40 ns, root-mean-square variation) and frequency stability ($1.45\times10^{-16}$ at 30 days, $8.8\times10^{-17}$ at 50 days), despite only 6% optical-clock availability. These results highlight a robust and realistic approach to immediately capitalize on the enhanced stability of the best optical clocks for international timekeeping. Moreover, our architecture is flexible and could further benefit from the addition of other optical clocks and stable laser oscillators [20-21] for better performance.

---

[*] The fractional frequency is a dimensionless quantity defined as $\left(\frac{measured\ frequency}{nominal\ frequency} - 1\right)$, which characterizes the frequency stability of a clock. In the following context, we may use "frequency" to stand for "fractional frequency" for simplicity, which can be easily identified from the unit.

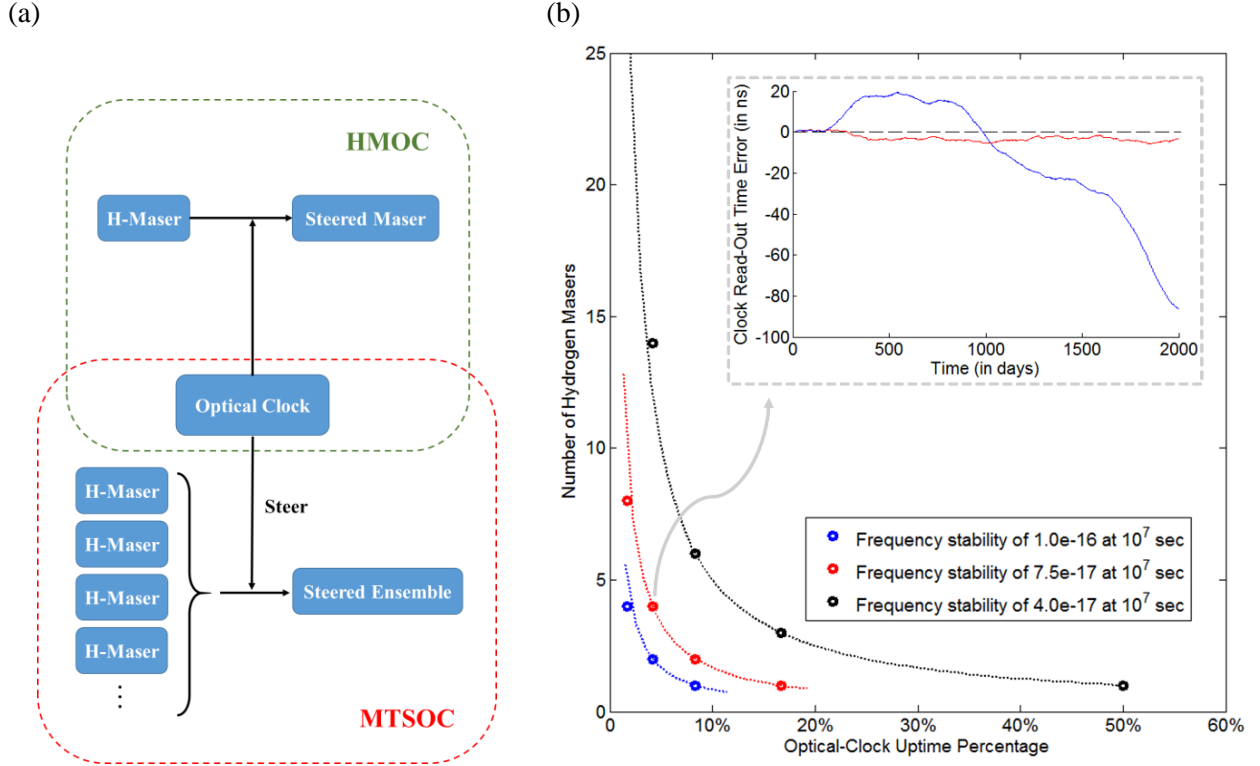

Figure 1. Concept of an optical-clock-based time scale. (a) illustrates the "Hydrogen maser + optical clock (HMOC)" architecture (green dashed box) and the "microwave time scale + optical clock (MTSOC)" architecture (red dashed box). (b) summarizes the relation between the number of masers and the optical-clock uptime for different performance goals (blue: $1.0\times10^{-16}$ at $10^7$ sec; red: $7.5\times10^{-17}$ at $10^7$ sec; black: $4.0\times10^{-17}$ at $10^7$ sec), when an optical clock runs once a day. The dots are the results of simulations, and the dashed curves are hyperbolas which well fit the dots. The inserted plot of (b) shows an example of the simulation, in time series. The blue solid curve is the read-out time error of a microwave time scale composed of four Hydrogen masers, and the red solid curve is the read-out time error of a MTSOC, composed of the four Hydrogen masers and an optical clock of 4.2% uptime. Note, (b) is plotted based on the simulations in [19].

## II.  Experimental Scheme of Optical-Clock-Based Time Scale

Figure 2 shows the experimental details of the optical-clock-based time scale (i.e., AT1' time scale) based on the MTSOC architecture. An Yb optical-lattice clock, which has a frequency of 518295836590863.71 Hz for $^1S_0 \rightarrow \ ^3P_0$ transition according to the latest worldwide comparison and weighted average [22], is used to stabilize a Ti:sapphire optical frequency comb [23]. The realization of the stabilization can be illustrated by Equation (1).

$$f_{laser} = f_0 + n \cdot f_{rep} + f_{beat}, \qquad (1)$$

where $f_{laser}$ is the frequency of the laser signal that is frequency-doubled and locked to the Yb atom's quantum transition. The comb's carrier offset frequency $f_0$ is locked to 10 MHz. $n$ is an integer. $f_{rep}$ is the frequency comb's repetition frequency which is in microwave region (close to 1 GHz). The beat frequency between the laser and the nearest comb tooth $f_{beat}$ is locked to 640 MHz.

By Equation (1), the Ti:sapphire frequency comb converts the Yb clock signal to the microwave region, via generating the repetition frequency $f_{rep}$. This microwave signal is compared with an up-converted Hydrogen maser signal at 1 GHz, via a mixer. The maser used in this experiment is labeled ST15 at NIST and provides the reference for all locked comb beatnotes. The frequency difference between the two microwave signals is measured by two frequency counters: a software-defined-radio counter [24], and a commercial frequency counter. The Yb clock is run in normal operation with a formal accounting of systematic clock shifts, including the gravitational redshift correction due to the height of the Yb clock from the geoid, altogether below $1\times10^{-17}$. Offsets in the frequency-comb-based optical-to-microwave synthesis were characterized and confirmed to be small ($< 3\times10^{-17}$), and the frequency counters exhibited negligible bias ($< 1\times10^{-17}$).

From the frequency counter data, we derive the fractional frequency difference between the Yb clock and the Hydrogen maser, i.e., $y_{Yb-HMaser}$. As proposed in Section 1, the MTSOC architecture is more favorable than the HMOC architecture, because the free-running microwave time scale is less noisy than the maser which allows more aggressive steering. We observe that the existing microwave time scale AT1 is better than a single Hydrogen maser by approximately a factor of 2. Another advantage of this architecture is that it is more reliable than the HMOC architecture because the time scale is only minimally affected by the failure of one of its contributing member clocks [9]. To achieve the MTSOC architecture, we need to know the fractional frequency difference between Yb and AT1, $y_{Yb-AT1}$. Because we continuously monitor the fractional frequency difference between the Hydrogen maser and AT1 (i.e., $y_{HMaser-AT1}$), we can calculate $y_{Yb-AT1}$ using the following equation.

$$y_{Yb-AT1} = y_{Yb-HMaser} + y_{HMaser-AT1}. \qquad (2)$$

With the information of $y_{Yb-AT1}$, we can correct the frequency and frequency-drift error in AT1 using a Kalman filter and thus generate the MTSOC time scale, AT1' time scale. The details of the Kalman-filter steering are discussed in Appendix B.

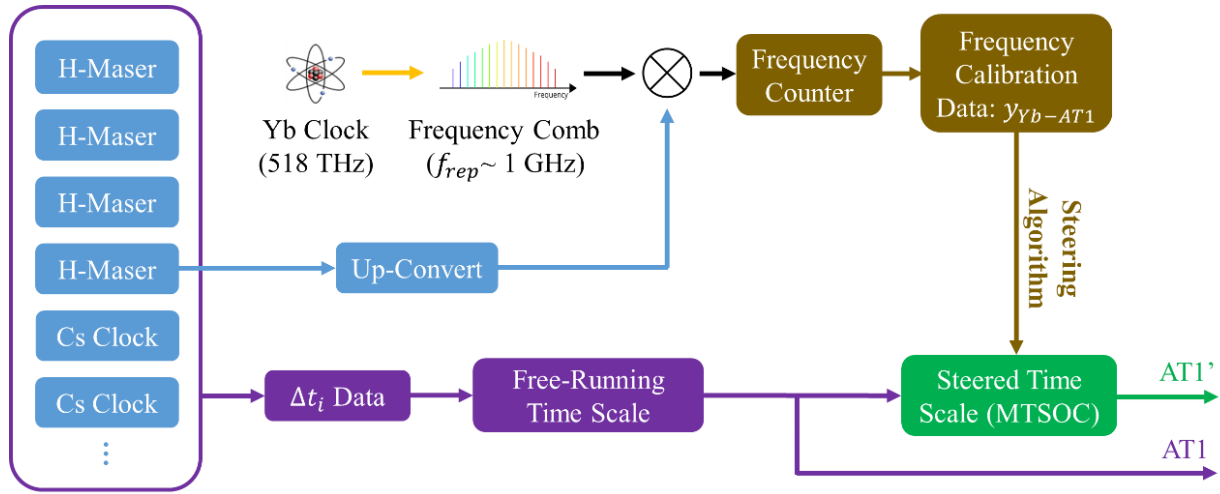

Figure 2. Experimental scheme of optical-clock-based time scale AT1', and comparison to free-running time scale AT1.

### III. Results of the Campaign in MJD 58054 – 58214

The Yb clock ran regularly during the period of late October 2017 – early April 2018 (MJD 58054 – 58214). Individual run times ranged from a few minutes to several hours, depending on the experimental arrangement. The total running time of the Yb clock was 241.8 hours. We obtain the average frequency difference $y_{Yb-AT1}$ for each run using Equation (2), the results of which are plotted in Figure 3(a). $y_{Yb-AT1}$ is around -4.26×10$^{-13}$, with a scatter of approximately ±5×10$^{-15}$. Note that this non-zero frequency difference comes from the frequency offset of AT1; as AT1 is a free-running time scale, no attempt is made to keep its frequency accurate.

The error bars in Figure 3(a) are assigned based on the frequency stability of "Yb – AT1" $\sigma_{Yb-AT1}$, which is determined in segments – less than 12 min, 12 min to 2 hours, and more than 2 hours. In general, shorter Yb clock runs result in larger uncertainty. Considering that AT1 is organized into 12-min grids [9], if the Yb clock runs for less than 12 min, $\sigma_{Yb-AT1}$ is essentially $\sigma_{Yb-HMaser}$ which can be calculated using the $y_{Yb-HMaser}$ data. $\sigma_{Yb-AT1}$ during 12 min to 2 hours can be calculated straightforwardly: with a long dataset of "Yb – AT1" (e.g., 6 hours), we are able to calculate the frequency stability of "Yb – AT1" up to 1/3 of the operation time; Multiple long datasets are used for the same calculation to improve the confidence. For the case of runs longer than 2 hours, this calculation does not work as we do not have datasets lasting many hours. Nevertheless, we know that $\sigma_{Yb-AT1}$ is composed of $\sigma_{Yb}$ which is negligible, $\sigma_{measurement}$ which becomes tiny after 2 hours based on the experimental data, and $\sigma_{AT1}$. Thus, we use $\sigma_{AT1}$ to determine the error bar for Yb clock runs > 2 hours.

Based on the $y_{Yb-AT1}$ result in Figure 3(a), we steer frequency- and frequency-drift parameters of AT1 to Yb immediately after each Yb run using the Kalman filter in Appendix B and generate the steered time scale AT1'. A new Yb-clock run typically earns a weight of 5% - 20%, depending on the duration of the new run and the elapsed interval since the last run. We derive the time difference between AT1' and AT1 by integrating the frequency- and frequency-drift steering record.

To evaluate the performance of AT1', we compare AT1' to UTC, via the chain of AT1' → AT1 → UTC(NIST) → UTC. Remember that the frequency stability of "AT1' – UTC" is composed of the frequency stabilities of AT1', UTC, and the chain. Thus, the frequency stability of "AT1' – UTC" gives the upper-limit stability of AT1'. In fact, the frequency stability of UTC should be excellent though not perfect, since it is the most accurate time in the world based on global Cs/Rb fountains. Also, the links AT1' → AT1 and AT1 → UTC(NIST) add negligible noise. Although the link UTC(NIST) → UTC suffers significant noise of ~4×10$^{-16}$ s/s over 5 days of averaging due to long baseline time-transfer methods [25-26], this link noise becomes negligible after ~ 10 days of averaging, since the time-transfer noise is dominated by flicker-phase process while the UTC/AT1' noise is mainly white-frequency or flicker-frequency process. Taking these points into account, the frequency stability of "AT1' – UTC" should be mainly composed of that of AT1', after 10 days. In other words, the comparison between AT1' and UTC via the above chain is a valid method of evaluating the long-term ($\gtrsim$ 10 days) performance of AT1'.

In Figure 3(b), red dots show the time difference between the optical-clock-based AT1' and UTC during the campaign (i.e., MJD 58054 – 58214). For comparison, blue dots show the time difference between AT1 and UTC with a constant frequency offset of +4.278×10$^{-13}$ s/s removed. We observe that AT1 walks away from UTC by ~16 ns over 160 days due to frequency drift in the free-running AT1. In contrast, AT1' exhibits no frequency offset, nor frequency drift, and maintains a time variation of 0.4 ns in root-mean-square (peak-to-peak variation: 1.4 ns) over the same period.

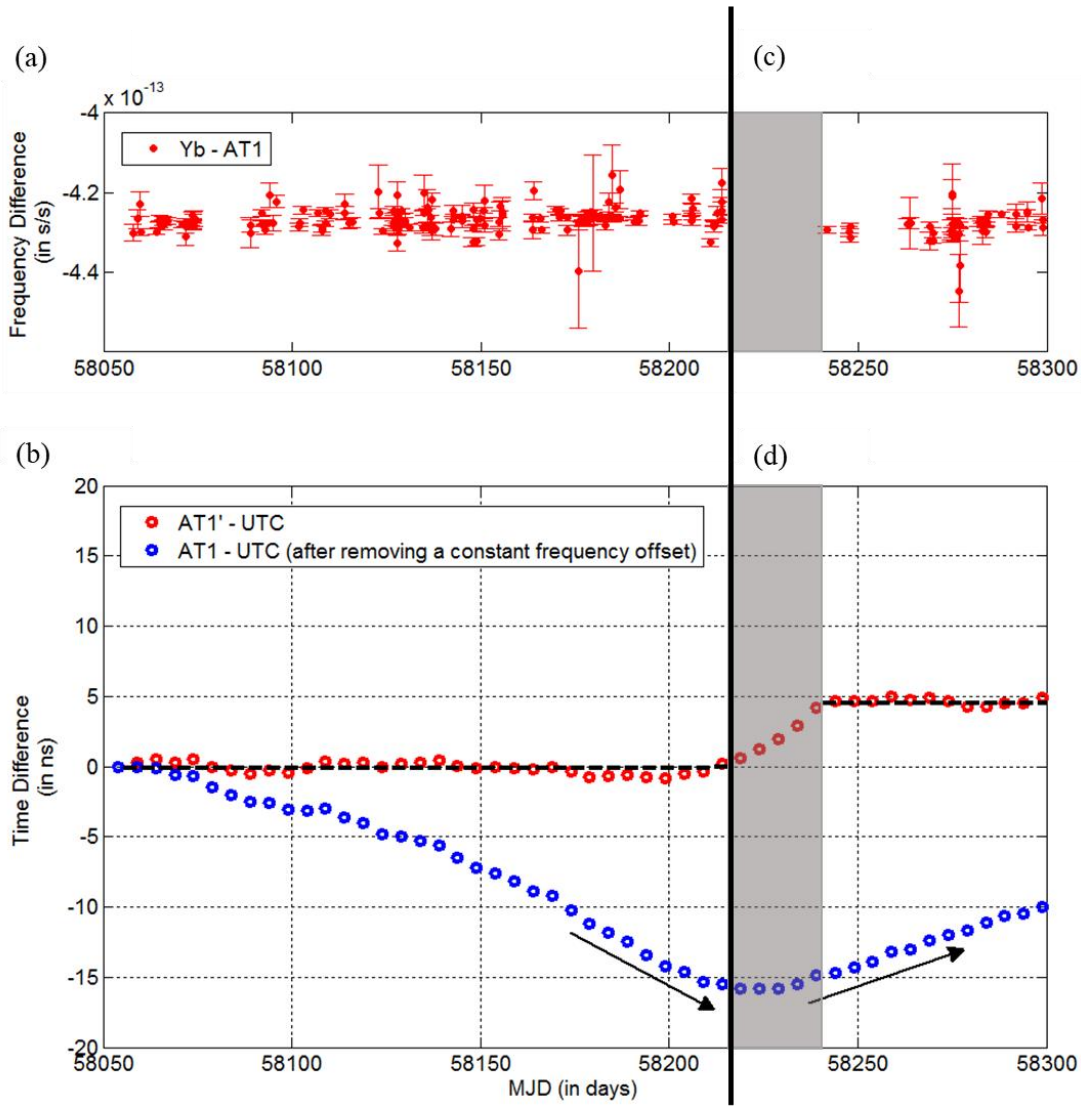

Figure 3. (a) shows the frequency difference between the Yb clock and AT1 during MJD 58054 - 58214. Note, AT1, composed of a few Hydrogen masers and a few commercial Cesium clocks, is a free-running microwave time scale at NIST. (b) shows the time difference between AT1' and BIPM UTC (red dots) during MJD 58054 - 58214. AT1' is the NIST time scale that is steered to the Yb clock. AT1' is set to 0 ns initially. AT1' has a root-mean-square variation of 0.4 ns with respect to UTC, during MJD 58054 – 58214. The time difference between AT1 and BIPM UTC (blue dots) is shown for reference. A constant frequency offset of $+4.278\times10^{-13}$ (measured from the first two points on the plot) in AT1 has already been removed. (c-d) show the behavior of AT1' and AT1 during MJD 58215 – 58300. During MJD 58215 – 58240 (grey region), the Yb clock ceased regular operation. After Yb-clock data resumed on MJD 58241, AT1' became flat with respect to UTC (black dashed line in (d)) indicating prompt frequency recalibration. The frequency change of AT1 is illustrated by the black arrows.

Besides the above time-series comparison between AT1' and AT1, we also compare frequency stabilities for this campaign (Figure 4). The stability of AT1' is comparable to that of AT1 for an averaging time of less than 20 days. After 20 days, AT1' significantly outperforms AT1. AT1' reaches $1.45\times10^{-16}$ at 30 days and $8.8\times10^{-17}$ at 50 days; in contrast, AT1 is $2.5\times10^{-16}$ at 30 days and $3.5\times10^{-16}$ at 50 days. The frequency stability of UTC(NIST) is shown for comparison. Because UTC(NIST) is usually steered to UTC weekly or bi-weekly for time accuracy, it exhibits worse short-term stability than AT1. However, this steering does accomplish better stability than AT1 over longer intervals. It is noteworthy that over all intervals AT1' – with just the benefit of Yb optical clock frequency calibrations – exhibits an improvement of a factor of 2 in the frequency stability over UTC(NIST) which has the benefit of UTC-informed time corrections.

The performance of existing Cs/Rb-fountain-based time scales, which are among the world's best, is also provided as a reference. At PTB (Physikalisch-Technische Bundesanstalt, Germany), during the same time period (i.e., MJD 58054 – 58214), two Cs fountain clocks [27] operated nearly 100% of the time. The time scale UTC(PTB) is composed of these two Cs fountain clocks and a few Hydrogen masers, and is gently steered to UTC in the long term for time accuracy. UTC(PTB) exhibits a peak-to-peak variation of 3.5 ns with respect to UTC, which is larger than that of AT1'. The frequency stability of UTC(PTB) is $2.5\times10^{-16}$ s/s at 30 days and $3.4\times10^{-16}$ s/s at 50 days (green curve in Figure 4). Similarly, at USNO (United States Naval Observatory), four Rb fountain clocks [28] (with performance comparable to Cs fountains) were running nearly 100% of the time. UTC(USNO), composed of the four Rb fountain clocks and dozens of Hydrogen masers, has a peak-to-peak variation of 3.5 ns with respect to UTC and a frequency stability of $3.8\times10^{-16}$ s/s at 30 days and $2.0\times10^{-16}$ s/s at 50 days (magenta curve in Figure 4).

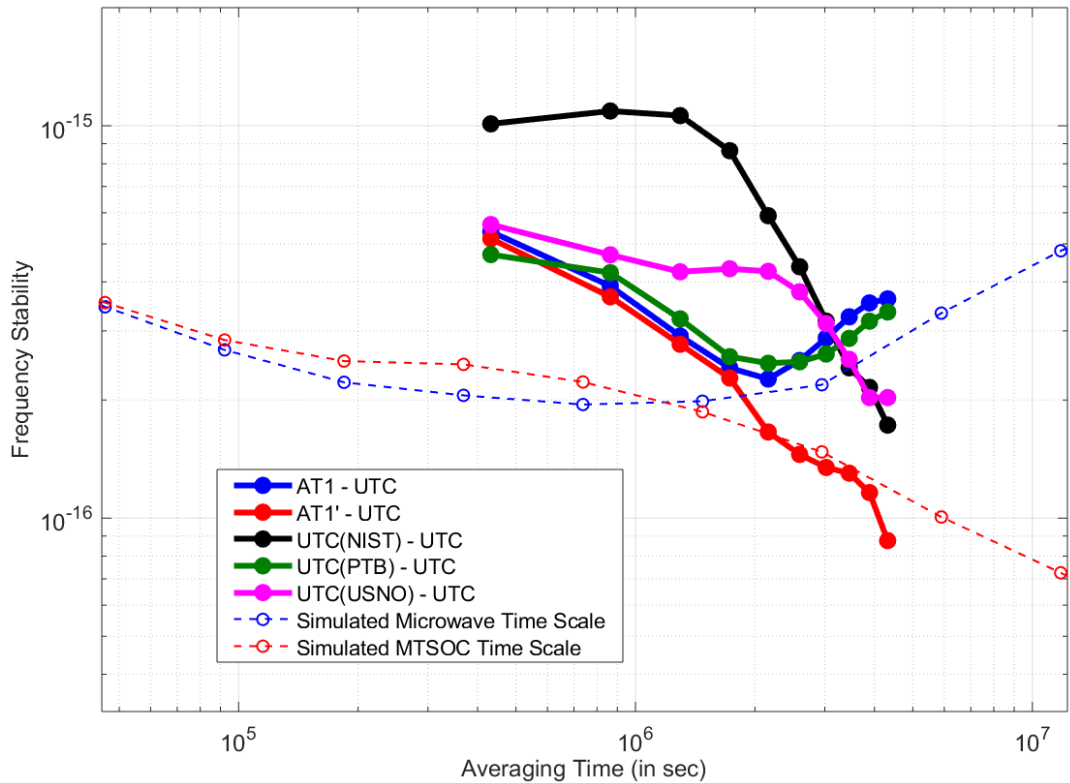

Figure 4. Frequency stability of AT1, AT1', UTC(NIST), UTC(PTB), and UTC(USNO), for MJD 58054 – 58214. The frequency stability is characterized by modified total deviation, and error bars are provided in Appendix C. The dashed curves show the simulation result. Note, the simulated MTSOC time scale (red dashed curve) is composed of the simulated microwave time scale (blue dashed curve) and a simulated optical clock that runs one hour per day.

# IV. Discussions

**IV(a). AT1' behavior in the absence of an optical clock**

Presently, there is no guarantee that an optical clock at NIST will operate once every few days. Here, we explore AT1' behavior when the optical clock stops running for an extended period and is subsequently re-introduced.

A gap in data from Apr. 07, 2018 (MJD 58215) – May 02, 2018 (MJD 58240) provides a good opportunity to explore this issue (see Figure 3(c)(d)). Intuitively, when an optical clock stops running, AT1' should degrade to AT1. However, since AT1' keeps its latest frequency/frequency-drift steering parameters, its exact time evolution can differ from AT1. Coincidentally, AT1 experienced a random positive frequency fluctuation during the data gap (annotated with black arrows in Figure 3). AT1', lacking Yb optical-clock calibrations, accumulated positive time offset accordingly (an error of ~ 5 ns). Upon resumption of Yb optical clock on MJD 58241, we see that AT1' quickly corrected its frequency and time error ceased accumulating. In the latter period of regular Yb optical-clock runs (MJD 58241 – 58300), AT1' remains a peak-to-peak time variation of 0.7 ns, consistent with the earlier behavior discussed in Section III.

Based on these observations, to maintain time- and frequency-accuracy in AT1', it is necessary that the optical clock reference run intermittently. We speculate that the required run schedule may be somewhat relaxed via use of a free-running timescale with better long-term stability (e.g., [29]).

**IV(b). Comparison between the real-data result and simulation**

Over the period of MJD 58054 – 58214, the Yb clock was operated and measured an average of 1.5 hours per day. Considering the non-uniform run schedule over that time (e.g., no weekend Yb-clock operation) which degrades the performance of the optical-clock-based time scale, the real situation is more comparable to a simulation of running an optical clock just one hour per day. The dashed curves in Figure 4 are the simulation result for which we assumed a noiseless reference time and no time-transfer noise. To be specific, the blue dashed curve is the simulated microwave time scale that represents AT1, while the red dashed curve is the simulated MTSOC result with an optical clock operating one hour per day.

By comparing the simulation result with the real-data result in Figure 4, we see that the real-data result is noisier up to ~ 10 days. We attribute this to the lack of time transfer and reference (i.e., UTC) noise in the simulations. The MTSOC architecture exhibits a significant improvement over the microwave time scale after 25 days in both simulation and real data. The divergence of the red curve from the blue curve in both simulation and real data indicates that the benefit of having an online calibration by an optical clock grows with time. Furthermore, both simulation and real data indicate that the MTSOC stability reaches the $10^{-17}$ regime after ~50 days. These observations generally validate the simulation results.

**IV(c). Future time keeping**

From both simulation and observed data, it is clear that a better flywheel oscillator (i.e., maser ensemble) offers an optical-clock-based time scale improved stability. Following this same logic, other types of flywheel oscillators under development may offer better performance still. For example, incorporating next-generation low-drift cavity-stabilized lasers [20-21] into the flywheel ensemble could yield a frequency stability $\sim 1 \times 10^{-17}$.

Alternatively, as noted earlier, higher optical-clock availability also offers time scale improvements. For example, with a consistent optical-clock availability at the 50% level, simulations indicate that our optical-clock-based time scale [19] could reach or exceed low $10^{-17}$ stability at ~100 days. While work is underway to develop an optical clock with high uptime, another path of increasing the optical-clock availability is to arrange for multiple optical clocks to contribute to AT1'. Indeed, as researchers explore the best optical clock candidates, many laboratories have developed more than one type of optical clock. Here at NIST, the Yb optical-lattice clock (featured above), the $Al^+$ trapped-ion quantum logic clock, and the JILA Sr optical-lattice clock could help calibrate AT1'. Even with only intermittent operation of each clock, a better long-term stability of AT1' results as the total optical-clock availability increases. This type of composite optical reference has the added advantage of redundancy: it is unlikely that multiple optical clocks would have long consecutive down times, avoiding the accumulation of large time error.

Beyond better local timekeeping, optical clocks offer improvements to International Atomic Time (TAI) and UTC [30]. Together with emerging time transfer techniques [31-35, 25], global time accuracy could reach the sub-nanosecond level. This level of performance has the potential to benefit a range of other technologies, including gravimetry, GNSS positioning, deep space navigation, and telecommunication. As pointed out in [36-37], the most accurate time constitutes a key element in the development of new gravitational measurement techniques. Remote comparisons of time could help a relativistic determination of gravitational potential. Sub-nanosecond time accuracy at the ground monitoring stations could enable the estimation of the GNSS satellite clocks' time offsets with a smaller uncertainty, reducing the error source that comes from satellite clocks in precise positioning [38]. In deep space cruise navigation, range and Doppler data are collected from a single Deep Space Network antenna over a pass typically lasting 8 hours or more, with the next pass several days to a week later [39]. A batch of passes over weeks to months are then used to solve for the trajectory. In this scenario, a sub-nanosecond ground time scale could help reduce the trajectory uncertainty coming from the reference-time uncertainty (note, another way of reducing the trajectory uncertainty is to develop a stable space clock as suggested in [39]). Finally, it is interesting to note that in telecommunication, the performance requirement of primary reference clocks has dropped from the ~$10^{-11}$ level of frequency uncertainty in 1988 (ITU-T G.811), to 100 ns in 2012 (ITU-T G.8272), and then to 30 ns in 2016 (ITU-T G.8272.1) [40]. Following this trend, one can expect this requirement to drop to a few ns in coming years, which can be challenging for the existing microwave time scales to support. In contrast, the performance of MTSOC as shown in this paper can fulfill this future requirement.

## V. Conclusion

This paper presents an optical-clock-based time scale (i.e., AT1') composed of an ensemble of continuously-operating microwave clocks and an intermittently-operating Yb optical clock. The new time scale, AT1', shows a root-mean-square variation of 0.40 ns with respect to UTC for more than 5 months; its frequency stability is $1.45 \times 10^{-16}$ at 30 days, and reaches the $10^{-17}$ regime at 50 days. This level of performance would allow national metrology institutes to provide more accurate time. With more optical clocks contributing to AT1', the availability of optical data can be increased, yielding even better long-term stability. To further improve the performance of AT1', we anticipate incorporating high-performance optical cavities into the clock ensemble.

**Appendix A: Dick effect limits on an optical-clock-based time scale**

The general problem of locking a noisy oscillator to a reference in the presence of dead time has been considered extensively in the atomic clock community, and the frequency stability limits derived from dead time are generally referred to as the Dick limit [41]. Armed with the noise spectrum of a local oscillator and the details of the dead time and measurement periodicity, it is straightforward to calculate the Dick limit, which is typically relevant at long times. Indeed, Dick effect considerations have been a key motivating force behind the development of improved optical local oscillators from cavity stabilization [42].

Here, we consider the Dick-limited frequency stability in the context of locking the frequency of a maser ensemble to that of an optical clock, when the optical clock is operated intermittently. We start with the simulated maser noise in [19]. To simply the clock model, we assume that the maser is composed of a white frequency noise of $1.26 \times 10^{-13}$ at 1 s, a flicker frequency noise of $3.09 \times 10^{-16}$ at 1 s, and a random walk frequency noise of $2.44 \times 10^{-19}$ at 1s. This model represents well the Hydrogen maser behavior after 10,000 s. Based on these parameters, we calculate the maser's frequency noise power spectrum. This noise spectrum is the primary input in the Dick effect calculation, which computes the aliasing of the maser noise into the optical clock, based on how frequently the measurements are made and with what dead time exists between measurements.

For the case of an optical clock being run once per day, the Dick limits are plotted by the dashed lines in Figure A1, for different scenarios. Note that the Dick effect results in a white frequency noise process which is plotted here for all times, but is really just relevant at longer times (e.g., $10^7$ s) where it would likely dominate the resulting frequency stability. We compare these Dick limits to the simulation results (solid curves), by focusing our attention at a sufficiently long time, e.g. $10^7$ seconds. We observe very close agreement between the two analyses, with the Dick limited frequency stability being within 90% of the simulation.

[19] points out that there is an improvement of $\sqrt{N}$ for an optical-clock-based time scale, by using an ensemble of $N$ masers rather than using a single maser. The Dick limit for the case of one maser and a four-maser ensemble are shown by the blue dashed line and the red dashed line in Figure A1, respectively. Comparing these two dashed lines, we find the improvement is 2, consistent with the square root of the number of masers. Both the simulation results and the Dick limits demonstrate the advantage of the "microwave time scale + optical clock (MTSOC)" architecture over the "Hydrogen maser + optical clock (HMOC)" architecture.

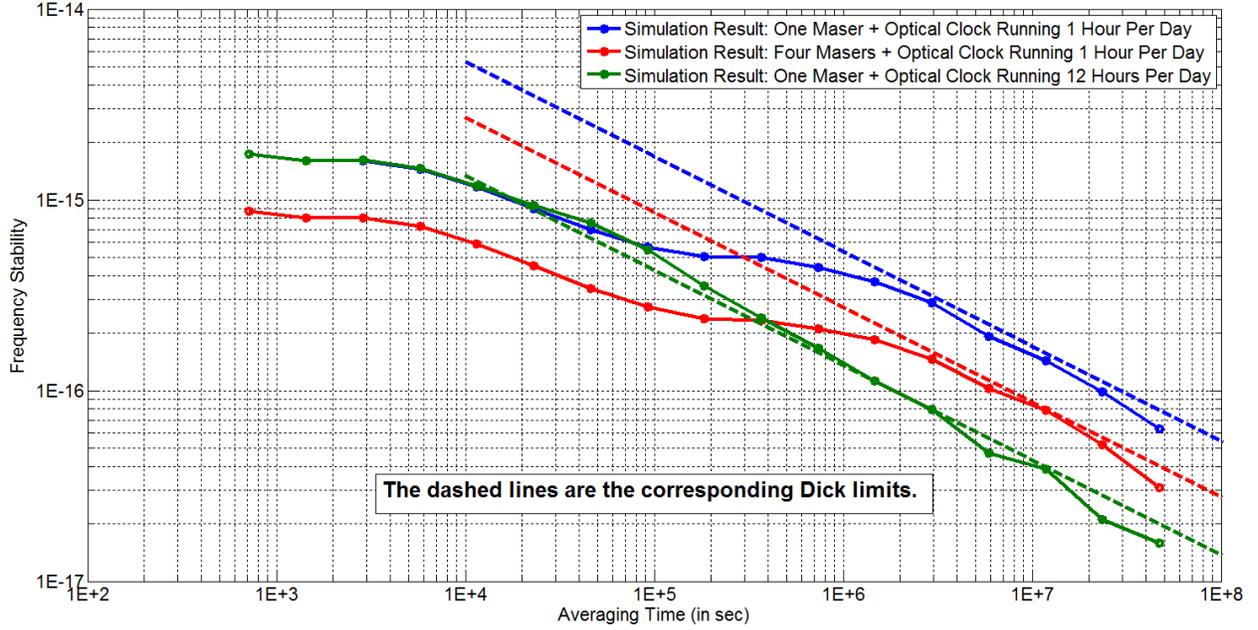

Figure A1. Comparison between the Dick limits (dashed lines) and simulation results in [19].

**Appendix B: Kalman-filter steering algorithm for optical-clock-based time scale**

A Kalman filter is used to estimate the frequency and frequency drift of the free-running time scale AT1 with respect to the Yb clock, and AT1 is steered based on this estimate [19, 43]. Here, we summarize the algorithm employed.

Equation (3) is the system model, which predicts the state of the system at epoch $k+1$ based on its state at epoch $k$. Here, $X(k)$ is the estimate state vector of the system at epoch $k$, and $X(k+1|k)$ is the predicted state vector of the system at epoch $k+1$. For our system, $X$ has two elements – the fractional frequency difference and the fractional frequency-drift difference between AT1 and the Yb clock. $\Phi$ is the transition matrix, which links $X(k)$ and $X(k+1|k)$. $u$ is the process noise, which is determined by the AT1 noise characteristics. Equation (4) is the measurement model. The $H$ matrix gives the relation between the state vector $X$ and the measurement vector $Z$. For our system, the measurement vector $Z$ is the average measured frequency difference between AT1 and the Yb clock during each operation period of the Yb clock and $H$ is (1  0). $v$ is the measurement noise.

$$X(k+1|k) = \Phi \cdot X(k) + u \tag{3}$$

$$Z(k+1) = H \cdot X(k+1|k) + v \tag{4}$$

According to the principle of Kalman filter, the estimated state vector at epoch $k+1$ $X(k+1)$ can be calculated using Equation (5).

$$X(k+1) = X(k+1|k) + K \cdot (Z(k+1) - H \cdot X(k+1|k)), \tag{5}$$

where $K$ is the Kalman gain matrix. From Equation (5), the estimated state vector is essentially a weighted average of the predicted state vector and the current measurement. The weight is determined by the Kalman gain matrix $K$. How to calculate $K$ is out of the scope of this paper and can be found in the classical book

[44]. Here, we want to address that if there is more than 15 days of missing optical-clock data, the adjustment operation is triggered. The reason why we choose 15 days as the threshold is that AT1 starts to exhibit a random walk process after 15 days and the Kalman filter cannot handle a random-walk process as well as a white-noise process. In the adjustment operation, we want to assign the weight of the prediction and the weight of the measurement based on their uncertainties. To be specific, the weight of the prediction should be proportional to $\frac{1}{\sigma_{AT1}^2(\tau=gap\ time)}$, while the weight of the measurement should be proportional to $\frac{1}{\sigma_{Yb-AT1}^2}$. Note, $\sigma_{Yb-AT1}$ is the frequency stability of "Yb – AT1" and Section III has discussed how to calculate it in details. Based on these weight calculations, we adjust the (1, 1) element of *K* accordingly. We emphasize that the adjustment operation is a one-time operation. As long as the optical-clock data gap is shorter than 15 days, *K* is still calculated based on [44].

An intuitive understanding of this filter is as follows: the longer the Yb clock runs, the larger weight the run gets; the longer time interval between the previous run and the current run, the larger weights the current run gets. Once we know the estimated state vector $X(k+1)$, we steer AT1 by adjusting its frequency and frequency drift and therefore generate AT1'.

## Appendix C: Error bars for the frequency stabilities in Figure 4

The data points in Figure 4 are computed using Stable 32 (Version 1.55), a popular scientific software for frequency stability analysis. In Figure 4, we choose the "all-tau" option in Stable 32, which provides the stability at every possible averaging-time value. Unfortunately, under the "all-tau" setting, Stable 32 does not provide the error bars. Here, we calculate the error bars for each data point in Figure 4, based on [45]. To be specific, according to Section 5.4.3 of [45], we can calculate the equivalent degrees of freedom for each data point. Then we use Equation (45) of [45] to get the corresponding confidence intervals (i.e., error bars). The 68.3% confidence level is used, to be consistent with the convention in the time and frequency community. Table A1 summarizes the results of the error-bar calculations.

Table A1. Error bars for the frequency stabilities in Figure 4. The unit for numbers in this table is $10^{-16}$. The 68.3% confidence level is used in this table, to be consistent with the convention.

|         | AT1 − UTC    | AT1' − UTC   | UTC(NIST) − UTC | UTC(PTB) − UTC | UTC(USNO) − UTC |
|---------|--------------|--------------|-----------------|----------------|-----------------|
| 5 days  | (4.82, 6.16) | (4.62, 5.91) | (8.90, 11.96)   | (4.14, 5.57)   | (4.97, 6.56)    |
| 10 days | (3.32, 4.96) | (3.16, 4.51) | (9.41, 13.44)   | (3.59, 5.36)   | (3.99, 5.96)    |
| 15 days | (2.40, 3.98) | (2.32, 3.64) | (8.89, 13.94)   | (2.70, 4.24)   | (3.50, 5.79)    |
| 20 days | (1.94, 3.53) | (1.87, 3.19) | (7.08, 12.10)   | (2.11, 3.61)   | (3.49, 6.32)    |
| 25 days | (1.78, 3.64) | (1.34, 2.43) | (4.76, 8.65)    | (1.95, 4.00)   | (3.42, 6.35)    |
| 30 days | (1.95, 4.38) | (1.15, 2.26) | (3.46, 6.81)    | (1.93, 4.35)   | (2.97, 6.01)    |
| 35 days | (2.20, 5.36) | (1.05, 2.25) | (2.48, 5.30)    | (2.00, 4.88)   | (2.43, 5.35)    |
| 40 days | (2.46, 6.53) | (1.00, 2.33) | (1.86, 4.34)    | (2.16, 5.75)   | (1.94, 4.69)    |
| 45 days | (2.64, 7.62) | (0.88, 2.24) | (1.63, 4.16)    | (2.37, 6.84)   | (1.53, 4.07)    |
| 50 days | (2.67, 8.49) | (0.66, 1.85) | (1.29, 3.67)    | (2.48, 7.86)   | (1.51, 4.45)    |

## Acknowledgments

We thank David Howe for the discussion on the frequency stability analysis. We also thank Neil Ashby, and Victor Zhang for their helpful inputs.

*Contribution of NIST – not subject to U.S. copyright.*